# Understanding the Earth as a whole system: From the Gaia Hypothesis to Thermodynamic Optimality and Human Societies

Axel Kleidon

**Abstract**

The notion that the whole is more than the sum of its parts has a long tradition in science. This, of course, also applies to the Earth system. With its myriad of processes, spanning from purely physical to life and human activity, the Earth is a vastly complex system. It may thus seem that there is nothing simple and general to say because of this overwhelming complexity. What I want to show here is that by formulating the Earth as a thermodynamic system, one can identify general directions and infer simple functioning because thermodynamics imposes fundamental limits on the dynamics. At the center of this description are energy conversions and states of disequilibrium, which are at the core of the dynamics of Earth system processes, from convection cells to living organisms and human societies. They are linked to each other and interact by their exchanges of energy and mass, and ultimately affect how much of the input of low entropy solar radiation from the Sun is converted into free energy, energy able to perform work, before the energy gets re-emitted by the Earth as high entropy terrestrial radiation. The emergent thermodynamic behavior of the Earth then becomes simple because the dynamics evolve to and operate at thermodynamic limits. Such behavior of Earth system processes operating at the edge of their limit can then be linked to previously described holistic theories, such as the Gaia hypothesis, with similarities in the described emergent behavior. Such a thermodynamic view, however, can go further, as it can also be used to understand the role of human societies in the Earth system and the potential pathways to a sustainable future. Thermodynamics taken together with the energy conversions and interactions within the Earth system can thus provide a basis to understand why the whole Earth system is more, and simpler, than the sum of its spheres.

# Introduction

The Earth reflects a planetary environment that is increasingly affected by human activity. The European landscapes reflect the imprints of the substantial human modifications over the last centuries, in which forests were cleared to increase agricultural production and towns and cities were built to provide homes for the growing population. This trend continues to date, with substantial rates of tropical deforestation, clearing highly diverse tropical rainforest ecosystems and replacing these with cash crops such as soybeans or oil palm plantations to benefit human societies. These transformations of land surfaces are accompanied by the trend of increasing rates of primary energy consumption, primarily derived from fossil fuels and essential in driving socioeconomic activity. The combustion of fossil fuels adds the waste product of carbon dioxide to the atmosphere, increasing greenhouse gas concentrations and causing global warming.

What may seem like a whole suite of individual problems, they share at a minimum one aspect: that their effects trickle down to the planetary scale, affecting the functioning of the whole Earth system. It raises questions about whether the growth of human societies is sustainable, or whether it has already exceeded the "*Limits to growth*" (Meadows et al., 1972), operating beyond so-called "*planetary boundaries*" (Rockström et al., 2009; Steffen et al., 2015) and it is just a matter of time before human societies collapse.

My goal here is not to go further into these questions, but rather shift the emphasis on the need to understand the Earth as a whole system. This understanding of the whole Earth, including the embedded role of human activity, should set the foundation to address these types of questions regarding global change and the sustainability of the human societies. Understanding the whole Earth may seem like a daunting, almost impossible task. Yet, if we can understand the generalities of how the Earth system with all its spheres works together as a whole system, we can better understand the consequences of human activity, anticipate how a sustainable future may look like, and whether *"planetary boundaries"* are cast in stone, or whether they can be pushed to higher levels, for instance by novel functioning introduced by human-made technology.

The aim of describing the whole Earth system, as opposed to a detailed description of its different parts, falls into a line of science that can be traced through much of the history of science. Already the Greek philosopher Aristoteles argued that the whole is more than the sum of its parts. This seems to suggest that there is more to learn by looking at the whole, and this should then also apply to the whole Earth. Alexander von Humboldt, the German natural scientist in the 18th century, aimed to describe this whole in his monumental book "*Der Kosmos*" (von Humboldt, 1845). Later, this aim continued with Josiah Gibbs, an American scientist with pioneering contributions to the development of thermodynamics in the 19th century, with his notion that the whole is simpler than the sum of its parts. His work on thermodynamics and its basis in statistical physics is an excellent example to demonstrate that the whole is more, and simpler, than its parts. Statistical physics aggregates the information of a vast number of molecules to a few macroscopic variables that describe a gas, in terms of its temperature, pressure, and density. It is able to predict how the gas as a whole — the collective sum of a myriad of colliding molecules — behaves when heat is added or pressure is changed, as, for instance, is reflected in the simple nature of the ideal gas law.

Yet, it is a far stretch from extending the notion of such emergent simplicity of a gas to the description of a whole planet. Thermodynamics is very well established as a physical theory, but how can we use it and gain similar insights for the whole Earth system? How can we use it to simplify the functioning of the Earth, illustrating that it acts simpler as a whole than the sum of its parts? And how can this view of the whole Earth inform us about the pressing questions regarding the sustainability of human societies?

In this chapter, I want to show that these questions can be answered when the whole Earth system is seen as a thermodynamic, interconnected system that not just includes the physical processes that shape climate, but also the biosphere as well as human activity (Kleidon, 2010; Kleidon, 2012; Kleidon, 2016). Central to this view is that any process on Earth essentially revolves around energy of different forms. Energy is what solar radiation provides as an input, it is kinetic energy that is reflected in planetary motion, organic biomass reflects chemical energy, and human societies need energy to sustain their activities. The different spheres of the Earth are connected by sequences of energy conversions (Figure 1). The energy input by the absorption of solar radiation heats the surface of the Earth, converting it into thermal energy. Some of this heat gets converted further to drive atmospheric convection, large-scale motion and the dynamics of the physical climate system. Another fraction of the solar energy gets taken up by photosynthesis to be converted into chemical energy in form of biomass, driving the dynamics of the biosphere. When harvested, it supplies the calories contained in the food to feed the human population. Energy thus trickles down sequences of conversions, changing its form as it is converted from the solar radiation to physical, biological and human forms of energy.

Another critical component is that these sequences of energy conversions do not just act in isolation from the top down to climate, the biosphere, and human societies. Processes alter the conditions under which these energy conversions take place, resulting in interactions and more complex behavior of the system (Figure 1b). The heat transported by planetary motion depletes the differences in warming due to solar radiation, thereby affecting how radiation is emitted back to space. The biosphere, by converting carbon dioxide, oxygen, water, and other resources, alters the chemical composition of the environment, thereby feeding back and altering climate, for instance by enhancing the continental cycling of water. And human societies, as already described, alter the system by transforming natural land into agricultural use and the atmospheric composition by combusting fossil fuels.

With these interactions, the Earth system appears even more complex. However, thermodynamics sets firm directions and limits to the conversions, and when these interactions are accounted for, these results in a new, emergent limit. I will argue that when sufficiently complex processes evolve to and operate at this limit, then the emergent behavior is simple, reflecting that the whole is more, and simpler, than the sum of its parts. The "*whole*" reflects the system as well as its interactions, and "*simple*" means that the emergent limit is predictive of the system's behavior. As all processes are connected by such sequences of energy conversions and interactions, ultimately to the input of solar radiation, it actually *requires* a description on the whole system to notice and derive such comparatively simple behavior of the Earth system as a whole.

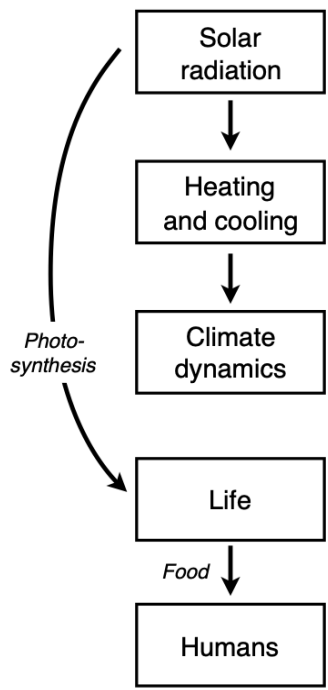
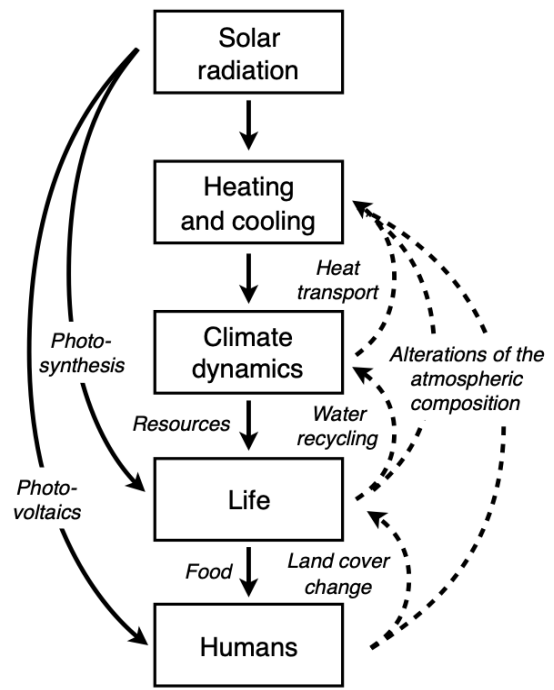

Figure 1: Illustration of (a.) a top-down view of energy conversions in the Earth system and (b.) the Earth as an energy-converting system with strong interactions.

#IMAGE1: Kleidon-Fig1-Earth.pdf

In the following, I first provide more context on thermodynamics and its application to life and the Earth system, and how these works have resulted in the formulation of the Gaia hypothesis, a hypothesis that states that life acts to alter Earth's conditions for its own benefit. I will derive a few criteria that describe this hypothesis as a means to capture its postulated behavior of the whole that I will use as a basis for comparison to the thermodynamic approach just described. I will then describe thermodynamics in more detail and how it applies to and constrains Earth system processes. I demonstrate this first by using the purely physical example of atmospheric motion, where this approach can explain the emergent characteristics of climate very well. I will then apply this approach to the role that life has in the Earth system, and continue with an outlook on human activity. I close with a brief summary and conclusions.

## Thermodynamics of Life and the Gaia Hypothesis

When it comes to life, the application of thermodynamics has a long history. Already Ludwig Boltzmann, the Austrian physicist from the 19th century who developed the statistical explanation of thermodynamics, explained that life draws its energy from the entropy difference of the hot Sun and the cold Earth (Boltzmann, 1886). Later, Erwin Schrödinger, an Austrian physicist with seminal contributions to quantum physics, described that the living cell follows the second law by consuming low entropy food and producing high entropy waste (Schrödinger, 1944). This notion of the living cell as a so-called dissipative system was picked up by James Lovelock and Lynn Margulis (Lovelock, 1972; Lovelock and Margulis, 1974) to describe that the Earth is just like a living cell, being a dissipative system. As living organisms are control systems that maintain homeostatic conditions inside their bodies, at least warm-blooded animals and humans, they argued that the Earth has similar regulatory mechanisms to keep conditions on Earth most suitable to life, resulting in homeostatic conditions on Earth. They called this hypothesis the Gaia hypothesis, alluding to the Greek goddess of Earth.

The Gaia hypothesis has been hugely influential in shaping the view of the Earth as a system. At the center of the proposed regulatory mechanism are the effects that life has on its planetary environment (Figure 2). While environmental conditions, such as the availability of light and water on land, affect photosynthesis and thus the level of biotic activity, these conditions are not independent of what life does. Tropical rainforests, for instance, are able to access soil water stored deep in the soil and transpire it back into the atmosphere (Nepstad et al., 1994). This allows rainforests to maintain transpiration through much of the dry season, thus returning more moisture to the atmosphere than a land surface without forests could do. This effect of dry season transpiration by rainforests contributes considerably to enhanced continental water recycling and precipitation on land (Kleidon and Heimann, 2000). As precipitation affects water availability, and water availability is a major limitation to biotic activity in tropical regions, it is not an independent forcing of biotic activity, but is at least in part affected by what terrestrial vegetation does. In doing so, rainforests tend to make their environmental conditions more favorable for productivity by making water availability less constraining. The effects of terrestrial vegetation form a positive feedback on productivity (or, a positive feedback on growth, Lenton 1998), with more vegetation being able to transpire more water, bring more vapor back to the atmosphere, enhance rainfall, thus enhancing water availability on land. A study with climate model simulations estimated that present-day vegetation acts to enhance continental rainfall by about 50% (Kleidon, 2002), so this effect is quite substantial. In other words, tropical rainforests seem to generate their own environment, at least in part, a notion that was already alluded to by Alexander von Humboldt in his *Ansichten der Natur* (1808).

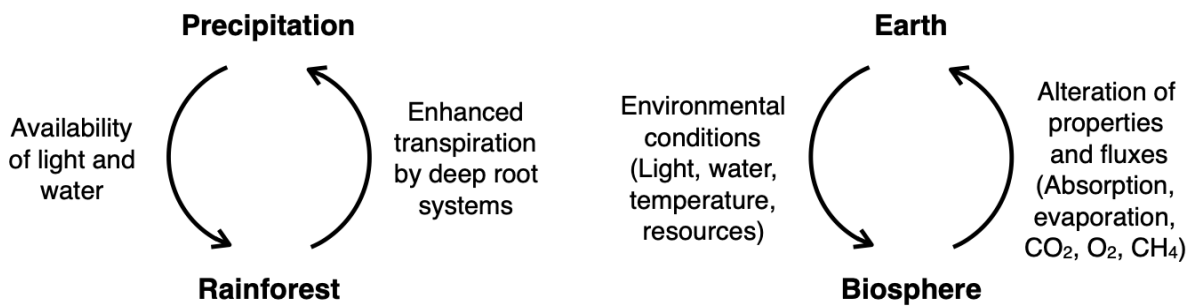

Figure 2: Interactions between life and the Earth system shown (a.) for the example of tropical rainforests and precipitation and (b.) in general play an important role in shaping the emergent conditions on Earth.

#IMAGE2: Kleidon-Fig2-Interactions.pdf

The transpiration of tropical rainforests is just one example of an effect that life has on the physical conditions of the Earth system. At the center of life is photosynthesis, the process which converts solar radiation into the chemical energy that sustains the metabolic reactions of the biosphere. To store this energy and build the biomass that forms living organisms, life needs to take up geochemical resources, carbon, water, and nutrients, from its environment. It thereby leaves an imprint and transforms the geochemical composition of the Earth system. These chemical transformations impacted the chemical composition of the atmosphere substantially over the Earth's history (e.g., Schwartzman and Volk, 1989), giving rise to a whole suite of biogeochemical feedbacks. These feedbacks that involve the activity of life on Earth are the central component of the regulatory mechanism that is proposed by the Gaia hypothesis.

Yet, unsurprisingly, the Gaia hypothesis is not without its critics. We only have the Earth as one example of a planet with life, so how can we generalize from its behavior to the role of life as a regulating entity and rigorously test this hypothesis? Geologists argued that the conditions on Earth have not been homeostatic as postulated by the Gaia hypothesis. And biologists argue that life has no goal function, and criticized the teleological, goal-seeking nature of the Gaia hypothesis. How would we know what is best for all life? What may be good for one species may be detrimental to another. A good summary of the criticisms of the Gaia hypothesis is summarized by Kirchner (1989).

Despite these valid criticisms, the Gaia hypothesis resembles outcomes that would match well the notion that the whole, Earth and life, is more, and simpler, than the sum of its parts. In the following, I want to show that this behavior can be explained at a general level when using thermodynamics in combination with an Earth system view that is based on energy conversions and interactions, as shown in Figure 1b. The central components of the Gaia hypothesis can be linked to four general criteria that are characteristic of thermodynamics and an Earth system view:

- **Disequilibrium.** Life represents a state of thermodynamic disequilibrium, allowing it to be a dissipative system. This is what stimulated Lovelock and Margulis to initially propose the Gaia hypothesis. Yet the notion of disequilibrium and dissipative system is not exclusive to living systems, but to a whole range of systems, from the physical climate system to human societies.
- **Interactions.** Life affects its environment, resulting in interactions. This notion was demonstrated above using the example of tropical rainforests and precipitation, and it is central for the regulatory mechanism proposed by the Gaia hypothesis for homeostatic conditions. Yet, effects and feedbacks are not exclusive to the biosphere either, but apply to purely physical climate dynamics and extend to human interactions with Earth.
- **Optimality.** The notion of the Gaia hypothesis of life regulating the Earth for its own benefit contains the implicit assumption of conditions being optimal for life. Yet, it raises the question as to what is being optimized. By formulating biotic activity in thermodynamic terms, one can use the dissipative activity as a goal function for the optimization, which, when optimized, implies the biosphere would be as active as possible. Yet, as physical processes and human activity are also dissipative, one would expect to find such optimizing behavior in these processes as well.
- **Homeostasis.** A central outcome postulated by the Gaia hypothesis is that the Earth is maintained in a homeostatic state. Such a homeostatic state implies the dominance of regulating, negative feedbacks, so that when the system is perturbed, it evolves back to its original state. When linking homeostatic behavior with optimality, it is a direct consequence of optimization. If a process is maximized, any perturbation would bring the process into a suboptimal state. Dynamics would then set in to bring the state back to the maximum state. In this sense, we would then expect such homeostatic behavior not just because of biotic activity, but in principle we would expect this outcome for any Earth system processes that evolves to and operates close to its optimum level.

As we will revisit these criteria further below, they are summarized in Table 1. Before I get to these, I will first provide some general basics on thermodynamics, how it applies to the Earth system, and explain the associated thermodynamic terms in more detail.

[Table 1] The four thermodynamic criteria associated with the proposed behavior by the Gaia hypothesis and how these apply to abiotic, biotic and human processes in the Earth system.

| Criterium | Atmosphere (Abiotic Earth system processes) | Biosphere (Biotic Earth system processes) | Anthroposphere (Human and socioeconomic processes) |
|---|---|---|---|
| **Disequilbrium** | Reflected in forms of free energy, such as kinetic energy in motion or potential energy of water stored at heights | Reflected in chemical free energy between carbohydrates, hydrocarbons, and oxygen | Reflected in human land cover maintained in the non-natural state, in human infrastructures and human-made technology |
| **Interactions** | Interactions result from heat transported by atmospheric motion that alters temperatures and radiative emission patterns of the planet. Hydrologic cycling alters radiative properties through clouds, ice, and water vapor. | Interactions result from biotic activity taking up resources from the environment, thereby altering the atmospheric composition and radiative properties of the atmosphere. Interactions by terrestrial vegetation result from greater surface absorption and evapotranspiration, thereby enhancing hydrologic cycling. | Interactions result from human-caused land cover change such as agriculture, as these change energy- and water balances, and reduce surface area available for the natural biosphere. Interactions due to primary energy consumption result from the depletion of stocks of chemical free energy stored in fossil fuels and changes in the atmospheric composition and radiative transfer. |
| **Optimality** | A state of maximum power associated with the conversion of heat into motion results from the coupling of motion with heat transport, which depletes temperature differences. This maximum power state is associated with optimum heat fluxes and temperature differences. | Associated with environmental conditions that maximize biotic productivity. This affects a primary limitation of biotic activity on land associated with the atmospheric exchange of carbon dioxide. | Associated with the depletion of stocks of free energy from the Earth system, such as biomass or wind energy, which impacts their renewal rates. Some forms of technology such as photovoltaics allow human societies to reach new levels of consumption as these make the Earth system more efficient in generating energy and resources. |
| **Homeostasis** | When a process evolves to and is maintained at a state of maximum power, its emergent state is inherently associated with stabilizing, negative feedbacks. Perturbations then cause dynamics that brings the process back to its optimum state. This results then in homeostatic-like behavior. | | |

# Thermodynamics of the Earth System

Thermodynamics is a physical theory that deals with energy conversions. It developed around the same time at which steam engines were invented at the onset of the industrial revolution in the 19th century and is nowadays a very well established theory. The physical basis of thermodynamics is extremely general and applies far beyond steam engines. Its relevance can in fact be seen in essentially all of what happens in the world around us.

To understand this very basic nature of thermodynamics, we need to describe Earth system processes in terms of the energy they convert. The primary source of energy for the Earth system comes in form of solar radiation. When absorbed, it turns into heat, or thermal energy, with some of it being further converted into the kinetic energy of atmospheric motion and the different forms of energies associated with hydrologic cycling. Similarly, plants convert some of the sunlight they absorb into the chemical energy associated with carbohydrates and produce the energy that fuels the biosphere as well as human societies by the food that they derive from it.

Thermodynamics sets two basic rules to these energy conversions. First, whenever energy is converted from one form into another, the different terms add up so that overall, energy is conserved. This is known as the first law of thermodynamics. The second law of thermodynamics deals with entropy, a measure for how energy is dispersed at the scale of atoms and molecules. When we look at this so-called microscopic scale, energy comes in discrete amounts called quanta (this notion led to the development of quantum physics at the beginning of the 20th century). These energy quanta can come in different forms. In form of radiation, the quanta are called photons and carry discrete amounts of radiative energy. In atoms and molecules, electrons can be associated with different shells, reflecting different intensities of being bound to the nucleus. These reflect different, discrete amounts of binding energies stored in the chemical bonds. Last, energy is stored in vibrations, rotations, and random motion of molecules, also in discrete amounts. Since energy is discretized and comes in these quanta, we can count them and associate probabilities of how these are distributed among the different states. Entropy is then simply the probability of this distribution.

The second law of thermodynamics then states that during energy conversions, the probability of distributing energy can, overall, only increase or stay the same. In other words, energy is being distributed in more and more probable ways, transforming energy to higher and higher entropy. This sets a profound direction for processes, which is why the second law has also been referred to as the arrow of time. When applied to the Earth system, we deal with three different forms of entropy, associated with radiation, electrons in chemical compounds, and molecules in solids, liquids, and gases. To evaluate the second law, we need to trace entropy and its conversions from the radiative exchange of the planet, with associated fluxes of radiative entropy, to heat, with associated thermal entropy, as well as the entropy associated with different chemical compounds.

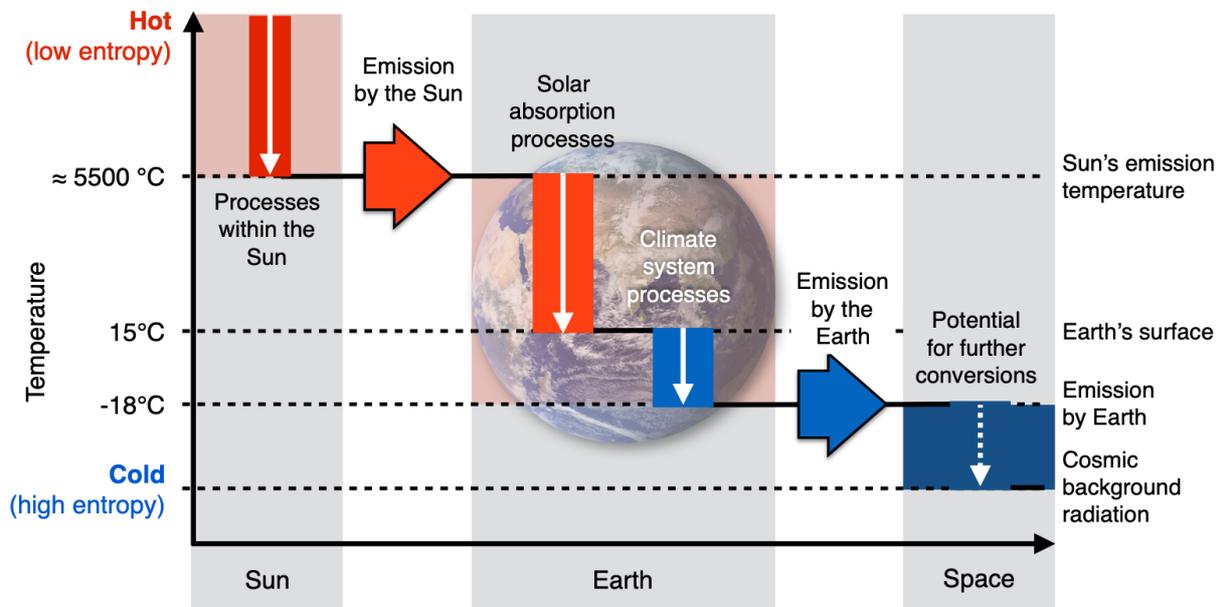

Figure 3: Energy is degraded to higher radiative entropy on its way from the Sun through the conversions within the Earth system to Space, which is a manifestation of the second law of thermodynamics.

#IMAGE3: Kleidon-Fig3-Degradation.pdf

To illustrate how entropy and the second law set the direction for Earth system processes, let us look at the changes of energy when solar radiation is absorbed by the Earth system, is converted into different forms, and is eventually re-emitted to space (Figure 3). When radiation is emitted from the hot Sun at about 5800K, it emits radiation with very low entropy. It is reflected in the radiative energy being distributed over comparatively few photons, with short wavelengths mostly in the range of visible light. Once absorbed by Earth and converted into thermal energy at the Earth's surface, the entropy has substantially increased because the energy is degraded to a much lower temperature of about 15°C or 288K. When this energy is eventually emitted to space, it is emitted at an even lower temperature of about -18°C or 255K, representing the so-called radiative temperature of the Earth. This radiative temperature is obtained by balancing the overall absorption of solar radiation by the Earth system with its emission according to the Stefan-Boltzmann law. The underlying assumption in this law is that the energy is emitted at maximum entropy, so that emitting the absorbed energy at the radiative temperature exports energy at the highest possible entropy to space. This higher entropy compared to solar radiation is reflected by the shift to less energetic radiation with longer wavelengths in the infrared range. It is associated with energy being distributed over many more photons (about 22 times as many), each of them being less energetic than those of solar radiation. Overall, the energy budget of the planet is closed, with absorption balanced by emission, but the emitted radiation exports a lot more entropy than is added to the system when solar radiation is absorbed. It reflects the direction imposed by the second law, degrading solar radiation from the hot temperature of about 5800K to a flux of terrestrial radiation representative of Earth's cold emission temperature of about 255K.

This difference in radiative entropy fluxes at the planetary scale provides conditions for substantial entropy production, energy conversions and dissipative processes to take place on Earth. The second law also provides a constraint to how much energy at best can be converted to generate free energy. Free energy is energy with no entropy associated with it, as it is the result of work being performed. It represents, for instance, the kinetic energy of atmospheric motion or the potential energy of cloud water droplets in the atmosphere. During the generation of free energy, the second law needs to be obeyed, which leads to the well-known Carnot limit. It requires that during an energy conversion process, at least as much entropy is removed as is being added.

We can see how the Carnot limit results from thermodynamics by taking a look at a power plant as an example of an energy conversion process. The power plant generates electric free energy from the heat of combustion. Combustion adds heat of low entropy to the plant, because energy is added at a high temperature. The waste heat of the conversion process, which we can see in form of the white clouds emerging from the cooling towers, exports heat at much lower temperature, so it represents energy of high entropy. The Carnot limit then sets the limit to how much electric free energy can at best be generated.

The Carnot limit follows directly from the first and second law of thermodynamics. The first law tells us that the heat released during combustion is balanced by the waste heat flux through the cooling towers and the energy contained in the generated electricity. So there is less energy leaving the system through the cooling towers than is being added, because some of it leaves the plant as electric energy. The second law requires that the cooling towers export entropy at least at the same rate as entropy is added by combustion. The Carnot limit is obtained when these two entropy exchanges are equal to each other. It results in a limit that is represented by the product of the heat flux multiplied by the difference in temperatures at which heat enters and leaves the power plant, divided by the combustion temperature. This latter ratio of temperatures is known as the Carnot efficiency, and is very well established as a theoretical limit.

In the Earth system, energy is converted as it is in a power plant. The surface of the Earth is heated by the absorption of solar radiation, while the atmosphere emits radiation to space. This takes place at different temperatures, with the surface being warmer than the atmosphere. This temperature difference is associated with a difference in entropy fluxes, so work can be derived from the heating, which generates motion.

Motion, in turn, lifts moisture to cooler layers in the atmosphere, bringing it to condensation. In other words, atmospheric motion performs work of lifting water, linking it to hydrologic cycling. What emerges from this description is that the Earth system is organized by sequences of energy conversions, from solar radiation to other forms. This hierarchy of energy conversion is shown in Figure 4. It illustrates the conversions from the input of solar radiative energy into thermal energy (or heat), kinetic energy, and further into potential and chemical forms of energy. So when we think of Earth system processes, the associated free energy is a key component. The magnitude of the dynamics is then set by how intensively these forms of energy are being generated, before they eventually get dissipated back to heat by friction.

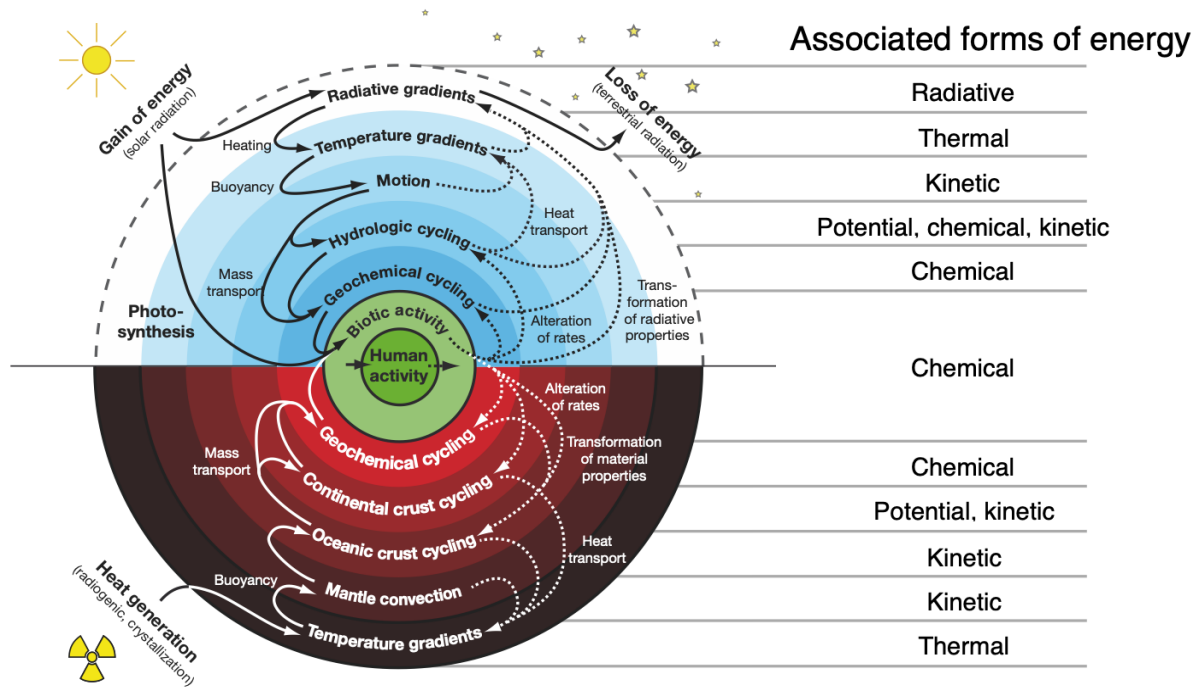

Figure 4: Solar radiation adds energy of low entropy to the Earth system, which is converted by different processes into different forms. The resulting dynamics distributes this energy, changes the material and radiative properties of the planet, so that the thermodynamic state of the planet results from these energy conversions as well as their interactions. After Kleidon 2010.

Yet, energy conversions in the Earth system are not a one-way street. When atmospheric motion is generated from heating differences, then this motion transports heat, affecting radiative gradients (shown by the dotted arrows in Figure 4). Hydrologic and geochemical cycling alter the chemical composition of the atmosphere, thereby affecting radiative transfer, so that they affect the radiative entropy exchange of the planet.

The emergent thermodynamic state of the Earth system is thus characterized by the intensity by which energy is being converted, yet the magnitude of these conversions is not just set by the solar energy input and thermodynamics. These effects are highly relevant, as they result in feedbacks that affect how much free energy can at best be generated by the planet.

This holistic picture of the Earth system will set the framework for the rest of this chapter. We will next go into more detail on the thermodynamic limits of energy conversions and look at how it plays out in the climate system that form the outer shells of Figure 4.

## Thermodynamics and the Atmosphere

So how do thermodynamic limits apply to the physical dynamics of the Earth system? To see its role, let us start by looking at what happens when sunlight heats the Earth's surface. The surface warms, raising its temperature and the emission of radiation from the surface. As the surface emits radiation, some of it gets absorbed within the atmosphere by greenhouse gases, and subsequently re-emitted, upwards, but also back down to the surface, resulting in the atmospheric greenhouse effect. These radiative fluxes form the central part of the surface energy balance, which budgets the energy fluxes that heat and cool the surface and that set its temperature. This surface temperature is warmer than the overlying atmosphere and the temperature at which radiation is ultimately emitted to space.

These two temperatures set up a temperature difference in the vertical that drives an atmospheric heat engine that generates the energy to sustain convective motion (Figure 5). The most power that this heat engine can generate is set by the Carnot limit, yet interactions with the surface further shape the maximum in power that can be generated. This is because convection transports heat away from the surface, in so-called sensible form and in form of moisture (called latent heat), both together referred to as turbulent heat fluxes. It results in additional cooling terms in the surface energy balance, lowering the surface temperature, and thus the efficiency term in the Carnot limit. Eventually, a maximum power limit is reached, setting an optimum heat flux and an intermediate temperature difference between the surface and the atmosphere. This trade-off is illustrated in Figure 5. It leads to the simple prediction that about half of the absorbed solar radiation is transferred from the surface to the atmosphere by convective motion, with the other half being transferred by the net exchange of terrestrial radiation between the surface and the atmosphere.

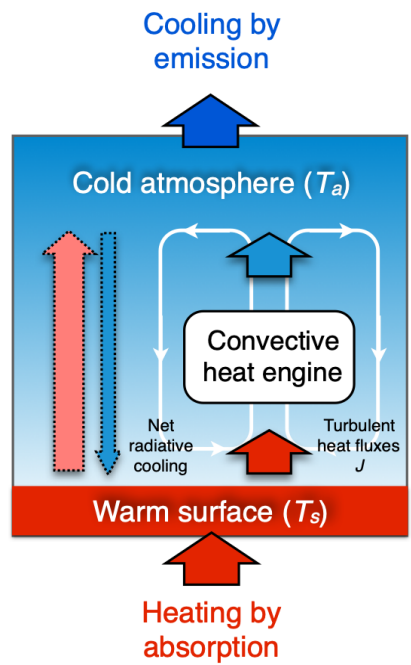
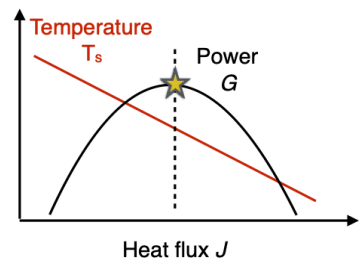

Figure 5: Description of atmospheric convection as a heat engine which converts some of the surface heating by solar radiation into kinetic energy. The maximum power that can be derived from the forcing is set by the thermodynamic Carnot limit as well as the interactions with the surface temperature, where more heat transport ($J$) by convection results in a cooler surface temperature ($T_s$), reducing the second term in the Carnot limit.

#IMAGE5: Kleidon-Fig5-MaxPower.pdf

The predictions of the maximum power limit work surprisingly well in a range of applications. It can predict the climatological partitioning of the surface energy balance reasonably well (Kleidon et al., 2014; Dhara et al., 2016), it can explain the observed diurnal variation in surface energy balance partitioning of different ecosystems (Kleidon and Renner, 2018), its response to tropical deforestation (Conte et al., 2019), the sensitivity of the hydrologic cycle to global change (Kleidon and Renner, 2013), and the difference in climate sensitivity between land and ocean (Kleidon and Renner, 2017). Some modifications may be necessary in the applications. For instance, in the application to the diurnal cycle, one needs to take into account diurnal heat storage variations within the atmosphere to derive the Carnot limit, as these variations represent not just variations in energy storage, but also in entropy changes (Kleidon and Renner, 2018).

Also, variations in the greenhouse effect need to be taken into account (Conte et al., 2019). Yet, overall, this range of application suggests that overall, atmospheric convection evolves to and operates at its thermodynamic limit. It works as hard as it can, maximizing the power to generate the kinetic energy associated with convective motion.

That this simple prediction by the maximum power limit works so well shows that the whole atmosphere is more, and simpler, than the sum of its parts. The "*whole*" stands here for the dissipative process of atmospheric convection, with its generation and dissipation of kinetic energy, but also includes the interactions of convection with the boundary conditions and the radiative forcing. This interaction is reflected by the heat transport intimately associated with convective motion that results in a lower surface temperature. This latter part is critical, as it represents the key tradeoff that shapes the maximum power limit and thereby sets the limit to how active atmospheric motion can be. The "*simple*" behavior originates here from this combination of thermodynamics with the interactions in an Earth system context.

What does this behavior of atmospheric convection have in common with the postulated behavior of the Gaia hypothesis? The emergent characteristics match the four criteria shown in Table 1 very well. First, atmospheric convection represents a process that is in thermodynamic disequilibrium. Power keeps generating the free energy in kinetic form associated with convective motion against friction, so it clearly is a case of a dissipative system. Second, interactions play a central role in setting the maximum power limit. It is the cooling effect of the heat transported by convective motion that results in the tradeoff and the maximum in power in the Carnot limit of the atmospheric heat engine. Third, this maximum power limit results in optimum behavior, reflected in an optimum heat flux and temperature difference associated with the maximum power limit. And last, it results in a homeostatic behavior. When the maximum power state is perturbed, convection will adjust to reach again the maximum power state. When the heat flux is below its optimum value, a larger temperature difference builds up that can generate more power, more motion, and thereby enhance the heat flux. Likewise, when the heat flux is above its optimum value, it will deplete the temperature difference, generating less power, which would slow down motion and reduce the heat flux. The power-maximizing atmosphere thus responds to perturbations with negative feedbacks. In other words, we have here a purely physical case in which we can find emergent behavior that essentially reflects the postulated behavior of the Gaia hypothesis.

Let's place this example back into the bigger picture of energy conversions in the Earth system that is shown in Figure 4. The example represents the outer shells of the figure in which radiative gradients drive temperature gradients and motion. Note that the same reasoning of maximum power can also be applied to the poleward heat transport that takes place at the planetary scale. The dashed lines in Figure 4 represent the heat fluxes that are associated with motion and that deplete the temperature differences in the Carnot limit. The maximum power limit thus reflects the most that the energy input from the Sun can be further converted into the inner shells of the figure. When analyzing the power involved, it represents less than 1% of the incoming solar radiation, reflecting a very low efficiency. This low efficiency is the result of the comparatively low temperature difference between the warm and cold reservoirs on Earth that is represented by the blue bar labeled "climate system processes" in Figure 3. Yet, atmospheric motion has its consequences on the overall thermodynamic state of the planet. Not only does it generate free energy and dynamics within the climate system, but by redistributing heat more evenly, it also acts to enhance the entropy export to space, making the Earth a more dissipative system.

## Thermodynamics and the Biosphere

As a next step, let us look at how thermodynamics applies to life in an Earth system context, so that we can evaluate it by the same four criteria. Applications of thermodynamics and a maximum power limit to ecosystems have a long tradition, starting with Lotka's work (1922a, b) and continuing with Odum and Pinkerton (1955) and Odum (1969). Yet, these applications are descriptive, and not quantitative and predictive as the examples of maximum power just described in the applications to the climate system. To be able to make it more quantitative, we would need to first identify the dominant bottlenecks for biotic activity, starting with a thermodynamic formulation of what biotic activity actually is.

To start, the activity of the biosphere can be described by the generation of free energy in chemical form by photosynthesis, and its dissipation by the metabolic activity of living organisms. The energy is generated when solar radiation is absorbed by chloroplasts, which split water, separate protons from electrons, drive a molecular machinery to form ATP, which in turn binds carbon dioxide and converts it into carbohydrates. In total, the generation of chemical free energy by photosynthesis converts the energy of 8 - 10 solar photons into the free energy contained in carbohydrates and oxygen, energy that is released into heat when these compounds are oxidized.

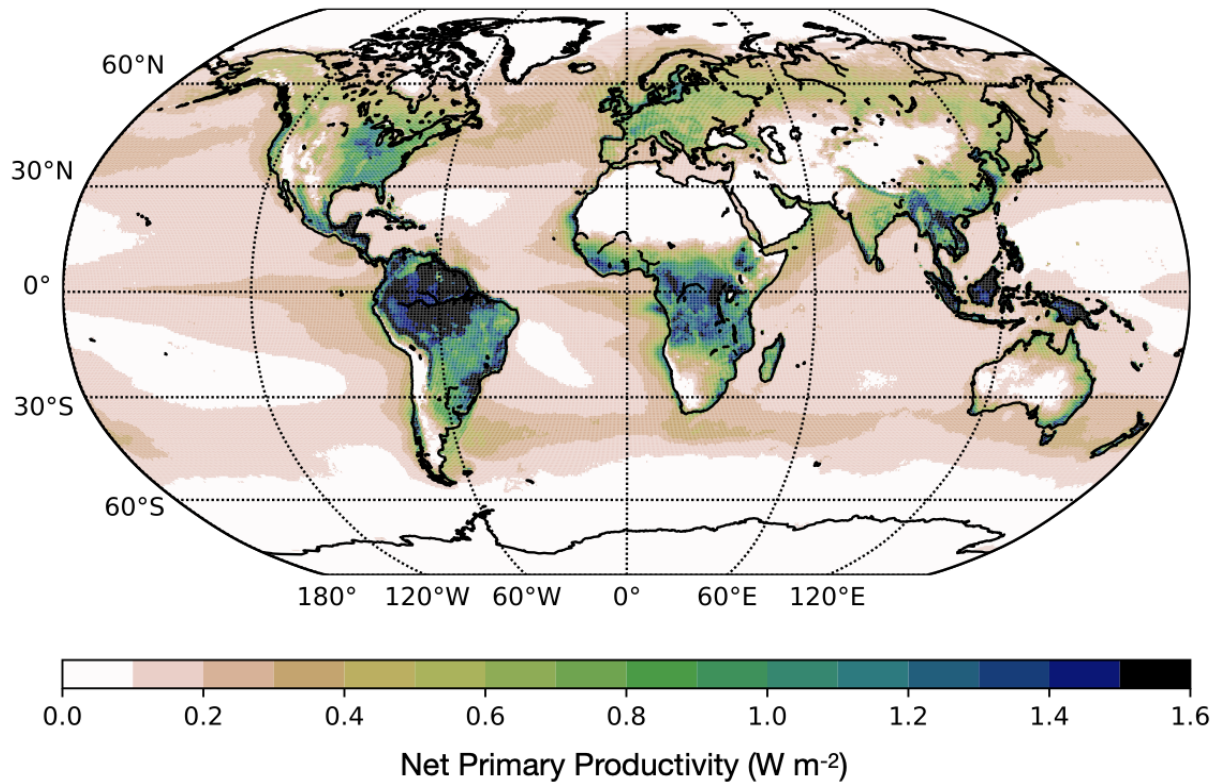

Figure 6: Net Primary Productivity (NPP) of the biosphere expressed in energy rates of W/m², estimated from satellite data as in Field et al. (1998). The data for the creation of this plot were obtained from the following websites: Ocean: http://orca.science.oregonstate.edu/2160.by.4320.monthly.hdf.vgpm.m.chl.m.sst.php. Land: https://nacp - files.nacarbon.org/nacp-kawa-01/

#IMAGE6: Kleidon-Fig6-Productivity.pdf

When we look at the efficiency of the photosynthetic conversion process, we find that with less than 3%, it is very low, even in well-watered and fertilized crop ecosystems (Monteith 1972, 1977). Only about half of the generated chemical energy turns into the biomass we observe, with the rate being the net primary productivity of the biosphere, setting the efficiencies to turn solar energy into biomass to typically 1% or less. A map of the distribution of this productivity, estimated from MODIS satellite data and converted into energy units, is shown in Figure 6.

This low efficiency of biotic activity is not what the direct application of thermodynamics to the conversion of radiation into chemical energy would tell us. When one looks purely at this conversion, noting that only about half of the solar spectrum can be utilized by photosynthesis, one would get an efficiency of about 17% (Hill and Rich, 1983). More detailed evaluations of the thermodynamics suggest maximum efficiencies of about 12% (Landsberg and Tonge, 1979), which is still notably higher than the observed maximum efficiency of < 3%.

So why is the efficiency of photosynthesis so low? We may get some hints for the reasons when we look at the spatial variation of biomass production shown in Figure 5, as these patterns reflect the dominant processes that constrain biotic activity. On land, this pattern is strongly shaped by water availability: Tropical regions with high rainfall and strong solar radiation show the highest rates of productivity on land, while in the subtropical desert regions, biotic productivity is absent. Biotic productivity on land actually needs a lot more water than what is needed during photosynthesis, because plants need to take up carbon dioxide from the atmosphere to store the energy harvest from sunlight in carbohydrates. In doing so, they inadvertently loose water to the atmosphere, a process called transpiration, that is, plant-mediated evaporation. In other words, plants basically trade water for carbon as they exchange material with the atmosphere. This trading of plants is well established in ecophysiological literature (e.g., Law et al., 2002), with the ratio of this exchange being referred to as the water use efficiency, with a typical value of about 3g of carbon dioxide gained for each kg of water lost. Water loss to the atmosphere, on the other hand, is a major part of the turbulent heat fluxes in the surface energy balance, and we have seen above that these are thermodynamically constrained. In other words, it appears that thermodynamics does not constrain biotic activity directly through the conversion from solar radiation to chemical energy, but rather indirectly by the material exchange of carbon dioxide between the vegetated surface and the atmosphere (Kleidon, 2016).

This notion of a thermodynamic control on the material exchange needed to sustain biotic activity is supported by the following, simple estimate. Of the 165 W/m$^2$ of solar radiation being absorbed on average at the Earth's surface (Stephens et al., 2012), about half of it is partitioned into turbulent heat fluxes according to the maximum power limit, of which evaporation plays the dominant role. A latent heat flux of about 82.5 W/m$^2$ corresponds to an evaporation rate of about 2.8 kgH$_2$O/m$^2$/day, using the latent heat of vaporization of about 2.5 MJ/kgH$_2$O for the conversion. With this evaporation rate, plants could take up about 8.4 gCO$_2$/m$^2$/day for photosynthesis, converting it into carbohydrates. As carbohydrates contain about 470 kJ of energy for each mole of carbon, or 10.7 kJ/gCO$_2$, this leads to an energy conversion rate of about 1 W/m$^2$. This is less than 1% of the absorbed solar radiation, supporting the low conversion efficiency of biotic activity.

We can make this notion of thermodynamic control of biotic activity plausible when we look at the canopies of tropical rainforests. Imagine what rainforests would do if it is the material exchange of carbon dioxide between the leaves and the atmosphere that is the bottleneck and limits productivity. One would expect that they would make the contact area through which this exchange takes place as large as possible, and ventilate it as much as they can. It would seem that this goal is exactly what the enormously heterogeneous canopies of rainforests with multiple layers of leaves achieve. This layout of rainforest canopies would seem well suited to maximize the surface area for material exchanges to alleviate this limitation.

For the marine biosphere, the notion of mixing as the primary bottleneck is well established. This is why the highest productivity in the open oceans are found in the mid-latitudes, where the synoptic activity of low pressure systems stirs the surface waters and provides greater degree of mixing. This mixing, in turn, brings nutrient-rich water from below to the surface where the productive phytoplankton resides. Greater mixing thus supports greater biotic productivity in the marine biosphere.

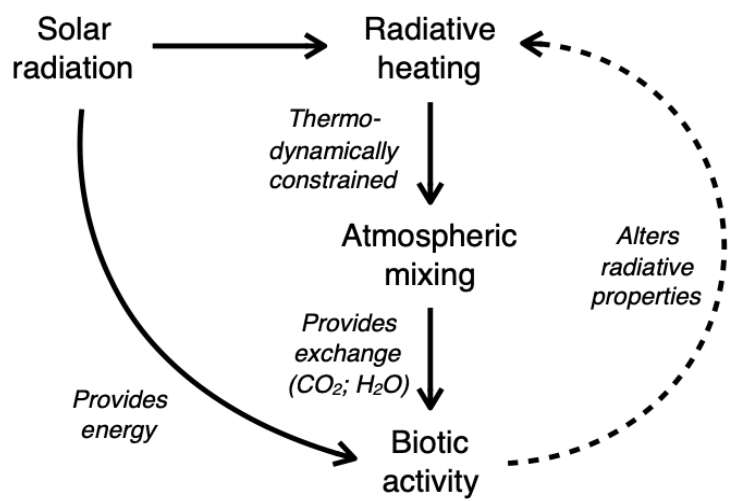

Figure 7: Biotic activity feeds back to planetary conditions by causing changes in the radiative properties (e.g., surface absorption, greenhouse gas composition) that affect atmospheric mixing and material exchange, thus impacting biotic activity.

#IMAGE7: Kleidon-Fig7-Feedback.pdf

When we place this notion of a thermodynamic exchange constraint on biotic activity back into the planetary picture shown in Figure 4, we notice how important and closely tied biotic activity becomes to the physical functioning of the climate system. The strength of climate dynamics does not simply play a role in terms of how much water it provides to land, but also in terms of the power to mix air between vegetated surfaces and the overlying atmosphere to provide carbon dioxide to the photosynthesizers or to stir the ocean to recover nutrients.

Biotic activity affects this power to mix through its effects on radiative properties. Rainforest canopies are not just highly heterogeneous, but they are typically also darker than non-vegetated surfaces, so they absorb more of the incoming sunlight and can drive more convection. At the planetary scale, biotic activity generates substantially more geochemical energy than abiotic processes (see Kleidon, 2016). This is because photosynthesis uses the low entropy energy contained in sunlight directly to do chemistry, tapping into the red bar labeled "solar absorption processes" in Figure 3. Abiotic processes are either based on the destructive use of ultraviolet radiation to drive ozone photochemistry, or derive it indirectly from heat engines, resulting in much lower magnitudes. Biotic activity is thus able to drive many more geochemical conversions at greater intensities, which alter the atmospheric composition. This affects the concentrations of greenhouse gases, the transfer of terrestrial radiation through the atmosphere, the maximum power limit to convection, which in turn affects how well the atmosphere is mixed. This causal link of processes provides the means for a positive feedback as shown in Figure 7 by which biotic activity can alleviate and minimize the constraint imposed by material exchange on biotic activity.

We could thus think of biotic activity on Earth as a mechanism to fine tune the atmospheric composition and the radiative properties of the Earth system such that the power to mix is further maximized. As geochemical reactions take place on slow time scales, it would probably take much longer to achieve a maximum power state than what it would take for atmospheric convection with its much faster dynamics and shorter time scales. The result of this, however, would be that biotic activity is maximized given the constraints of mass exchange. And as this mass exchange constraint varies at regional scales with climate, it results in biotic productivity that can be predicted from climate variables by relatively simple means, as reflected in Figure 5.

It would, again, support the notion that the Earth's biosphere as a whole is more, and simpler, than the sum of all living organisms. The "*whole*" refers in this case to the biosphere, consisting of all life on Earth, but also to the physical constraints of the Earth system that limit its activity as well as to the interactions that alter these constraints. After all, the level of biotic activity does not take place in isolation (as it is shown in Figure 1a), but it is intimately tied to the geochemical conversions and cycling within the Earth system. It is constrained by the material exchange associated with motion, which in turn is thermodynamically limited. The "*simple*" refers here to the level of predictability of the emergent patterns of overall biotic activity that shows clear geographic variations with climate (Figure 5). No matter which species are present and what they do, the aggregated activity of these ecosystems show clear and simple geographic variations that reflect the dominant climatological constraints, which were linked here to the material exchange and water availability. It is indicative of the activity of the biosphere to evolve to and operate at this limit set by the physical environment. This does not imply teleology, which was one of the objections raised against the Gaia hypothesis. It simply reflects the dynamics of what one would expect from complex thermodynamic processes taking place in the Earth system context, just like atmospheric convection.

Let us next link this picture of biotic activity back to the Gaia hypothesis and evaluate the four criteria from Table 1. First, biotic activity is clearly a dissipative process which maintains the biosphere in a state of thermodynamic disequilibrium. It generates chemical free energy by photosynthesis, the disequilibrium is represented in chemical form by the simultaneous presence of carbohydrates and oxygen, and metabolic activities of plants and animals deplete this disequilibrium to make a living. Defining biotic activity by its generation and dissipation of chemical free energy allows us to quantify in a clear and consistent way how active the biosphere is, compare it to the activity of purely physical processes, and evaluate it in terms of which environmental effects are beneficial or detrimental to life. The effects of biotic activity give rise to interactions between the physical conditions and the activity of the biosphere and involve the effects on radiative properties and water availability (see also above when the Gaia hypothesis was introduced, Figure 2). These interactions can result in a positive outcome, making environmental conditions more suitable for biotic activity up to a point when mass exchange or resource availability such as water constrains productivity. Optimality is reached when biotic activity evolved to a level at which it is only constrained by environmental conditions. At this optimal state, perturbations would again cause negative feedbacks to play out to bring the activity of the biosphere back to its maximum. This, in turn, would be similar to the homeostatic conditions for and by the biosphere that was postulated by the Gaia hypothesis.

So it would seem that the Gaia hypothesis is not as far-fetched as it may initially seem. Yet, when interpreting Earth's history, we likely need to consider that the coevolution of the biosphere with the geochemical and physical conditions of the Earth took millions of years. The past conditions of much of Earth's history were then probably more shaped by the overarching trend towards greater biotic activity towards the maximum rather than reflective of an optimum biosphere that regulates the Earth system.

## Thermodynamics and the Anthroposphere

At the end, let us look at human activity in the same way and see whether it can also show such Gaian-like behavior. Human activity, at its very core, involves energy conversions. The human species, playing the role of a consumer in the Earth's biosphere, needs to draw its chemical free energy from its environment to sustain its metabolic needs. This energy is contained in the calories we eat as food, which we draw from the products of photosynthesis in form of crops and other agricultural products. Looked at in this way, human activity becomes a thermodynamic process embedded tightly in the Earth system, taking chemical free energy from the environment that was generated by the biosphere and dissipating it into heat and carbon dioxide.

The energy consumed by the metabolic activity of human societies can be estimated from previous studies that evaluated how much of the productivity of the biosphere is appropriated to human use (e.g., Vitousek et al., 1986; Haberl et al., 2007). These studies repeatedly find that this appropriation represents a sizable fraction of the net primary productivity on land. When we convert this carbon flux to a rate of chemical free energy consumption, it yields a magnitude of about $8 \times 10^{12}$ W (Kleidon, 2016). This rate of energy consumption can then be compared to other Earth system processes in terms of their dissipative activity. Photosynthesis of the Earth's biosphere, for comparison, produces about $220 \times 10^{12}$ W of chemical free energy. While human consumption is less than this number -- and it needs to be, because plants also need something to live on -- it is nevertheless of the same magnitude of Terawatt, or $10^{12}$ W. The large-scale motion in the atmosphere involves about $1000 \times 10^{12}$ W, again larger, yet being in a similar range. These energetic considerations support the notion that human activity is, by its magnitude, a significant thermodynamic Earth system process, affecting the state of the Earth system. This notion has already been captured by the conjecture that we have entered the geologic era of the Anthropocene (Crutzen, 2002).

Humans consume even more energy than just by their metabolic activities. Human societies use primary energy, mostly in the form of fossil fuels, to drive their socioeconomic activities. This energy is, again, a central component of what keeps societies economically active (e.g, Ostwald, 1909; Ayres and Nair, 1984; Fischer-Kowalski and Haberl, 1998), reflected in a tight correlation between primary energy consumption and gross domestic product, as an indicator of economic activity. It is the energy used for manufacturing, for transportation, heating and cooling buildings, lighting, and information processing by computers and the internet. It can be seen as an "outsourcing" of human activity to technology, which helps to make more things faster. A tractor plows a field faster than an ox or a human, and a container ship transports more goods and faster than carriages drawn by horses. Yet, it comes at a huge energetic cost. Human primary energy consumption in 2017 averaged to about $18 \times 10^{12}$ W, representing more energy than what is needed to meet the human metabolic demands. Consequently, the activity of the human sphere, or the Anthroposphere, consisting of the dissipative activity of human organisms as well as the primary energy consumed by their technology, is of even greater magnitude when compared to other Earth system processes.

A comparison to a Gaian-like behavior and the four criteria is then hampered by the fact that the current levels of energy consumption by human societies are not sustainable, particularly regarding their use of primary energy. This is because this consumption relies heavily on the depletion of chemical free energy stored in geologic deposits of fossil fuels that were built up by past biotic activity over millions of years of Earth history. This unsustainable state depletes the Earth's disequilibrium state in form of the hydrocarbons in geologic deposits and atmospheric oxygen, increases the concentration of greenhouse gases in the atmosphere, and causing the well-documented trends in recent decades of global warming. This, again, shares similarities to the previous examples of climate and the biosphere in that human activity affects the radiative properties of the atmosphere, resulting in feedbacks between human activity and planetary energy conversions depicted in Figure 4.

We can nevertheless imagine how a sustainable human future may look like. The term "sustainable" here gets a physical meaning: the ability to meet human energy demands in a steady-state, and not in a transient state that is maintained by the depletion of a stock of free energy that eventually will run out. We may take this further and even look for ways to meet human energy needs that are beneficial to the Earth system, with beneficial meaning that non-human Earth system processes may become thermodynamically more active. How would these theoretical considerations play out into something more concrete?

Let us first illustrate a few examples for optimum states that would allow for maximum sustainable levels of human activity in an Earth system context. If we consider an individual human being in the natural environment, one can imagine a level of maximum activity resulting from the simultaneous needs to acquire food and to cool the body (Kleidon 2009). The human metabolism consumes substantial amounts of energy, releasing about 80-100 W at rest, and up to 1200 W when physically active (for comparison, the Earth's surface absorbs about 165 W/m$^2$ of solar radiation on average). This heat needs to be given off to the environment. In a colder environment, productivity is typically lower yielding less food, but it is easier to give off the metabolic heat. In a warmer environment, these two factors shift, with potentially greater productivity, but reduced ability to give off heat. This may result in intermediate climates which allow for maximum levels of human activity. However, nowadays humans live mostly in engineered environments maintained by technology and trade food, so that these restrictions no longer apply.

Another maximum state can be demonstrated for the human appropriation of net primary productivity from the natural biosphere (Kleidon 2007). This human appropriation is associated with food production, but also includes, e.g., the use of wood as building material or as a primary energy source. The productivity of the biosphere depends on environmental constraints (as discussed above and shown in Figure 6), but also on the standing biomass. A mature forest is more productive than a group of seedlings because the forest is able to access more resources and absorb light more effectively. The more biomass humans appropriate, the less standing biomass is left behind, eventually reducing the productivity. This results in a maximum sustainable level of appropriation of biotic productivity that would set an upper limit to how much food can maximally be generated by natural means.

A similar maximum can be found for using wind energy from the atmosphere to generate renewable energy as a supply for primary energy (Miller et al., 2011; Miller and Kleidon, 2016). With more and more wind turbines deployed at large scales, these would take out more and more of the kinetic energy of the atmosphere, causing wind speeds to decline. If wind turbines were to be deployed over all land surfaces, these would at most yield about twice to three times the current primary energy demand. This would, again, set a maximum sustainable level for how much primary energy can be generated.

These two cases of appropriating energy from the Earth system both leave less energy behind, impacting the associated Earth system processes. Appropriating biomass results in a less active biosphere, because of reduced productivity and less chemical energy being left over to feed natural food webs. Using wind energy leaves less strong winds behind, weakening associated dynamics that are driven by frictional dissipation near the surface, such as generating ocean waves and currents, and the power to mix the upper parts of the oceans. In other words, these two cases, while being in principle sustainable, result in a weaker Earth system that is less thermodynamically active. It would result overall in detrimental effects of human activity on the Earth system, as shown in Figure 8a.

This notion of natural limits led to the notion of "*Limits to growth*" of Meadows et al. (1972), as mentioned in the introduction, a notion that can also result in societal breakdowns, as e.g., shown in Frank et al. (2018). However, this notion neglects the critical role of human-made technology, with some of this technology being better at tasks than the natural counterparts. With such technologies, human activity can contribute to generating more free energy in the Earth system and convert it more efficiently into resources, thus strengthening the Earth system (Figure 8b).

A prime example of such technology is energy generation by photovoltaics. Photovoltaics generates electric energy directly from the absorption of sunlight, without the intermediate conversion into heat (which is linked to the red bar labeled "Solar absorption processes" in Figure 3). Current, industrial-grade solar panels, with an efficiency of about 20%, are already vastly more efficient than the biological counterpart of photosynthesis, with an efficiency of less than 3%, or physical processes that are able to use less than 1% of the incoming solar radiation. Solar panels are so much more efficient because they are not heat engines, and they are better than photosynthesis, because the generated free energy is exported and distributed in electric form, not requiring mass exchange of carbon dioxide and water. The theoretical limit to this solar energy conversion is even higher at more than 70%, depending on technology, providing ample possibilities for future technological improvements. Humans thus contribute novel technological means to the Earth system that are, in principle, able to convert much more of the solar energy into free energy than what life or the physical climate system can accomplish. It is a basis for an evolution towards a more active Earth system that generates and dissipates more free energy than the pre-human, natural Earth.

An example for how technology can improve resource availability compared to the natural counterpart is desalination technology. The natural hydrologic cycle desalinates seawater by evaporation and subsequent condensation, which requires substantial amounts of heat that is supplied by the absorbed solar radiation. Human-made membranes can desalinate seawater with substantially less energy. This technology can enhance the hydrologic cycle at a lower energetic cost than nature, which can make more water available on land, can reduce water limitations, and can enhance terrestrial productivity. Imagine if this technology were to be used to green deserts. It would allow for a future expansion of agriculture into deserts, arid regions that are currently unproductive, avoiding the need to clear natural rainforests in the tropics to meet future food demands.

This would clearly lead to a different thermodynamic state of the Earth system. It would result in a state that is not only shaped by physical and biological dynamics, but also by dynamics sustained by human-made technology. It would be associated with greater energy throughput and more active material cycling. It would result in a planet with greater levels of sustained food production and human primary energy consumption, yet with the ability to minimize the impact on the natural biosphere. The whole Earth system would then also be greater than the sum of its parts, because the additional energy input by human-made technology would play a central role in driving additional dynamics and feedbacks that shape the emergent state.

To close, let us try and summarize this view of human activity and link it back to the four criteria shown in Table 1. First, human activity represents a dissipative process that is maintained in thermodynamic disequilibrium, just like any form of life, and like any physical dissipative process. To sustain human activity, energy consumption is central, to sustain the human metabolism as well as the socioeconomic activity. This disequilibrium is reflected in the organic carbon stored in our bodies and in the infrastructures and technology that we build. Using this energy consumption allows us to quantify the magnitude of human activity and we can compare it to other dissipative processes of the Earth system. Second, interactions associated with human activity play an increasing role and alter the Earth system. These interactions emerge because humans need to draw their energy from the Earth system, thereby immediately causing environmental impacts. As human activity has reached a magnitude comparable to other Earth system processes, its impact can be felt at the planetary scale, with altered greenhouse gas concentrations in the atmosphere and substantial parts of the continental surfaces put into human use.

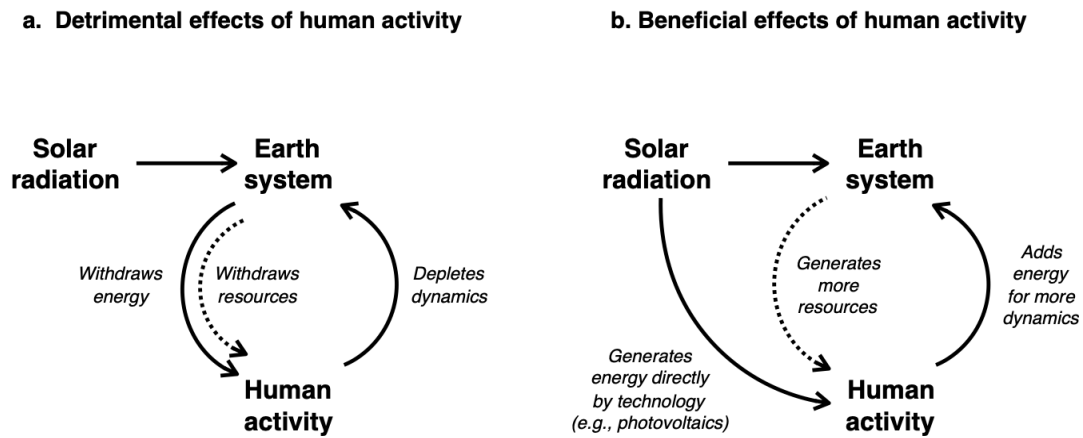

Figure 8: Human activity as an Earth system process that either (a.) consumes free energy from the Earth system, thereby weakening its dynamics, or (b.) generates additional free energy by technology and thereby strengthens the dynamics of the Earth system.

#IMAGE8: Kleidon-Fig8-Humans.pdf

It would seem that the last two criteria, optimality and homeostasis, do not quite apply to human activity at this point, but may become relevant once human societies transitioned into a sustainable state. We described a few cases of maximum levels that could result in similar dynamics as in the biosphere, yet through the addition of technology, such as photovoltaics and seawater desalination, this would drive an entirely different kind of evolution as it would bring the Earth system to unprecedented levels of dissipative activity. Such a human-dominated Earth would possibly also exhibit maximum levels of human activity and, possibly, associated homeostatic behavior, but this would of course be hypothetical at this point and would depend on how human societies make their decisions to reach a sustainable future.

## Synthesis

To close this chapter, let us briefly recap why the whole Earth system is more, and simpler, than the sum of its parts, based on this thermodynamic view of the Earth system. I described a view of the Earth system in which its different processes from the absorption of solar radiation to physical motion, biologic activity and human societies are quantified by the energy that they convert. These energy conversions reflect the laws of thermodynamics. While individual forms of free energy and associated states of disequilibrium are generated, for instance kinetic energy in the atmosphere or chemical energy in the biosphere, overall no energy is lost or gained.

Additionally, the second law sets the overall direction for these conversions towards higher entropy and it imposes thermodynamic limits that constrain the intensity of the dynamics.

Applied to Earth's atmosphere, this limit taken together with interactions with the boundary conditions results in a limit of maximum power that can predict observed climate and its response to change very well. The simplicity emerges here because i) the whole system is described, that is, atmospheric dynamics as well as its interactions with the boundary conditions set by radiative fluxes and temperatures, and because ii) the dynamics of the atmosphere are so complex that the dominant limitation is set by the thermodynamic maximum power limit, so that this limit can be used to describe the emergent behavior of the system.

A similar view was applied to the activity of the biosphere. Biotic activity is closely integrated with the thermodynamic functioning of the Earth system, not because of the limitations by sunlight, but rather by the constraints of mass exchange by carbon dioxide and water that are linked to the maximum power constraint of atmospheric motion and that shape patterns of productivity on land. Once this connection is being made, it shares similar features as atmospheric motion. The interactions are more subtle through changes in radiative properties, and the evolution of the dynamics to a maximum power state of the biosphere may proceed much more slowly and may rather be reflected as an evolutionary trend. I also described how the resulting behavior is, in fact, very similar to the proposed behavior by the Gaia hypothesis of Lovelock and Margulis.

Yet, the thermodynamic approach has more to offer, because it can also be applied to human societies. Here, the link to the Earth system is accomplished by the need of human societies for energy and resources. Given its magnitude of energy dissipation, human activity affects planetary functioning. It shares the same thermodynamic features as the physical and biological processes of the Earth system. One distinguishing characteristic of humans is their technology, which can achieve things more efficiently than their natural counterparts. I gave the two technologies of photovoltaics and seawater desalination as examples. With these technologies, it would seem that human societies have the means to sustainably grow further in the future beyond the limits to growth that would be set by the natural Earth system. These technologies would allow for increased socio-economic activity, energy consumption and material cycling, making the whole planet thermodynamically more active.

In conclusion, it would seem that the Earth system is more and simpler than the sum of its parts. The absorbed solar energy does not just simply flow and get transformed through the different spheres of the Earth in an arbitrary direction or in a straightforward, top-down manner. Interactions by the processes that transform the solar energy feed back to the radiative conditions of the planet, evolve to states of greater energy conversions and accelerate the second law to the maximum possible extent, given the thermodynamic constraints. This, in turn, results in simple, predictive behavior. The dynamics are then so complex that they reflect their thermodynamic constraints, so the emergent outcome can be predicted by relatively simple means.

# References

Ayres, R. U. and I. Nair. 1984. "Thermodynamics and economics." *Phys. Today* 37: 62–71.

Boltzmann, L. 1886. "Der zweite Hauptsatz der mechanischen Wärmetheorie." *Almanach der kaiserlichen Akademie der Wissenschaften* 36: 225–259.

Conte, L., Renner, M., Brando, P., dos Santos, C. O., Silvério, D., Kolle, O., Trumbore, S. E., and A. Kleidon. 2019. "Effects of Tropical Deforestation on Surface Energy Balance Partitioning in Southeastern Amazonia Estimated from Maximum Convective Power." *Geophysical Research Letters* 46: 4396-4403.

Crutzen, P. J. 2002. "Geology of Mankind." *Nature* 415: 23.

Dhara, C., Renner, M., and A. Kleidon. 2016. "Broad Climatological Variation of Surface Energy Balance Partitioning Across Land and Ocean Predicted from the Maximum Power Limit." *Geophysical Research Letters* 43: 7686-7693.

Field, C. B., Behrenfeld, M. J., Randerson, J. T., and P. Falkowski. 1998. "Primary Production of the Biosphere: Integrating Terrestrial and Oceanic Components." *Science* 281: 237–240.

Fischer-Kowalski, M., and H. Haberl. 1998. "Sustainable Development: Socio-economic Metabolism and the Colonization of Nature." *Int. Soc. Sci. J.* 50: 573–587.

Frank, A., Carroll-Nellenbeck, J., Alberti, M., and A. Kleidon. 2018. "The Anthropocene Generalized: Evolution of Exo-Civilizations and Their Planetary Feedback." *Astrobiology* 18: 503-518.

Haberl, H., Erb, K. H., Krausmann, F., Gaube, V., Bondeau, A., Pluttzar, C., Gingrich, S., Lucht, W., and M. Fischer-Kowalski. 2007. "Quantifying and mapping the human appropriation of net primary productivity in Earth's terrestrial ecosystems." *Annu. Rev. Environ. Resour.* 39: 363-391.

Hill, R., and P. R. Rich. 1983. "A Physical Interpretation for the Natural Photosynthetic Process." *Proc. Natl. Acad. Sci. USA* 80: 978-982.

von Humboldt, A. 1808. *Ansichten der Natur.* Tübingen: J. G. Cotta.

von Humboldt, A. 1845. *Kosmos. Entwurf einer physischen Weltbeschreibung.* Stuttgart & Tübingen: J. G. Cotta.

Kirchner, J. W. 1989. "The Gaia Hypothesis: Can It Be Tested?" *Review of Geophysics* 27: 223-235.

Kleidon, A. 2002. "Testing the Effect of Life on Earth's Functioning: How Gaian is the Earth System?" *Climatic Change* 66: 271-319.

Kleidon, A. 2006. "The Climate Sensitivity to Human Appropriation of Vegetation Productivity and its Thermodynamic Characterization." *Global and Planetary Change* 54: 109-127.

Kleidon, A. 2009. "Climatic Constraints on Maximum Levels of Human Metabolic Activity and Their Relation to Human Evolution and Global Change." *Climatic Change* 95: 405-431.

Kleidon, A. 2010. "Life, Hierarchy, and the Thermodynamic Machinery of Planet Earth." *Phys. Life Rev.* 7: 424–460.

Kleidon, A. 2012. "How Does the Earth System Generate and Maintain Thermodynamic Disequilibrium and What Does it Imply for the Future of the Planet?" *Phil. Trans. R. Soc. A* 370: 1012–1040.

Kleidon, A. 2016. *Thermodynamic Foundations of the Earth System.* Cambridge, UK: Cambridge University Press.

Kleidon, A., and M. Heimann. "Assessing the Role of Deep Rooted Vegetation in the Climate System with Model Simulations: Mechanism, Comparison to Observations and Implications for Amazonian Deforestation." *Clim. Dyn.* 16: 183-199.

Kleidon, A., and M. Renner. 2013. "A Simple Explanation for the Sensitivity of the Hydrologic Cycle to Surface Temperature and Solar Radiation and its Implications for Global Climate Change." *Earth System Dynamics* 4: 455-465.

Kleidon, A., and M. Renner. 2017. "An Explanation for the Different Climate Sensitivities of Land and Ocean Surfaces based on the Diurnal Cycle." *Earth System Dynamics* 8: 849-864.

Kleidon, A., and M. Renner. 2018. "Diurnal Land Surface Energy Balance Partitioning Estimated from the Thermodynamic Limit of a Cold Heat Engine." *Earth System Dynamics* 9: 1127-1140.

Kleidon, A., Renner, M., and P. Porada. 2014. "Estimates of the Climatological Land Surface Energy and Water Balance Derived from Maximum Convective Power." *Hydrology and Earth System Sciences* 18: 2201-2218.

Landsberg, P. T., and G. Tonge. 1979. "Thermodynamics of the conversion of diluted radiation." *J. Phys. A* 12: 551–562.

Law, B. E., Falge, E., Gu, L., Baldocchi, D. D., Bakwin, P., Berbigier, P., Davis, K., Dolman, A. J., Falk, M., Fuentes, J. D., Goldstein, A., Granier, A., Grelle, A., Hollinger, D., Janssens, I. A., Jarvis, P., Jensen, N. O., Katul, G., Malhi, Y., Matteucci, G., Meyers, T., Monson, R., Munger, W., Oechel, W., Olson, R., Pilegaard, K. U., Paw, K. T., Thorgeirsson, H., Valentini, R., Verman, S., Vesala, T., Wilson, K., and S. Wofsy. 2002. "Environmental Controls over Carbon Dioxide and Water Vapor Exchange of Terrestrial Vegetation." *Agric. For. Meteor.* 113: 97–120.

Lenton, T. M. 1998. "Gaia and natural selection." *Nature* 394: 439–447.

Lotka, A. J. 1922a. "Contribution to the Energetics of Evolution." *Proc. Natl. Acad. Sci. USA* 8: 147–151.

Lotka, A. J. 1922b. "Natural Selection as a Physical Principle." *Proc. Natl. Acad. Sci. USA* 8: 151–154.

Lovelock, J. E. 1972. *Gaia: A New Look at Life on Earth*. Oxford: Oxford University Press.

Lovelock, J. E., and L. Margulis. 1974. "Atmospheric Homeostasis by and for the Biosphere: the Gaia Hypothesis." *Tellus* 26: 2–10.

Meadows, D. H., Meadows, D. L., Randers, J., and W. W. Behrens III. 1972. *The Limits to Growth*. New York: Universe Books.

Miller, L. M., Gans, F., and A. Kleidon. 2011. "Estimating Maximum Global Land Surface Wind Power Extractability and Associated Climatic Consequences." *Earth Syst. Dynam.* 2: 1–12.

Miller, L. M., and A. Kleidon. 2016. "Wind Speed Reductions by Large-scale Wind Turbine Deployments Lower Turbine Efficiencies and set low Generation Limits." *Proc. Natl. Acad. Sci. USA* 113: 13570-13575.

Monteith, J. L. 1972. "Solar Radiation and Productivity in Tropical Ecosystems." *J. Appl. Ecol.* 9: 747-766.

Monteith, J. L. 1977. "Climate and the Efficiency of Crop Production in Britain." *Phil. Trans. R. Soc. B* 281: 277–294.

Nepstad, D. C., de Carvalho, C. R., Davidson, E. A., Jipp, P. H., Lefebvre, P. A., Negreiros, H. G., da Silva, E. D., Stone, T. A., Trumbore, S. E., and S. Vieira. 1994. "The Role of Deep Roots in the Hydrological and Carbon Cycles of Amazon Forests and Pastures." *Nature* 372: 666–669.

Odum, E. P. 1969. "The Strategy of Ecosystem Development." *Science* 164: 262–270.

Odum, H. T., and R. C. Pinkerton. 1955. "Time's Speed Regulator: The Optimum Efficiency for Maximum Power Output in Physical and Biological Systems." *Am. Sci.* 43: 331–343.

Ostwald, W. 1909. *Energetische Grundlagen der Kulturwissenschaften*. Leipzig: Klinkhardt.

Rockström, J., Steffen, W., Noone, K., Persson, A., Chapin, F. S., Lambin, E. F., Lenton, T. M., Scheffer, M., Folke, C., Schellnhuber, H. J., Nykvist, B., de Wit, C. A., Hughes, T., van der Leeuw, S., Rodhe, H., Sörlin, S., Snyder, P. K., Constanza, R., Svedin, U., Falkenmark, M., Karlberg, L., Corell, R. W., Fabry, V. J., Hansen, J., Walker, B., Liverman, D., Richardson, K., Crutzen, P., and J. A. Foley. "A Safe Operating Space for Humanity." *Nature* 461: 472–475.

Schrödinger, E. 1944. *What is Life? The Physical Aspect of the Living Cell*. Cambridge, UK: Cambridge University Press.

Schwartzman, D. W., and T. Volk. 1989. "Biotic Enhancement of Weathering and the Habitability of Earth." *Nature* 340: 457–460.

Steffen, W., Richardson, K., Rockström, J., Cornell, S. E., Fetzer, I., Bennet, E. M., Biggs, R., Carpenter, S. R., de Vries, W., de Wit, C. A., Folke, C., Gerten, D., Heinke, J., Mace, G. M., Persson, L. M., Ramanathan, C., Reyers, B., and S. Sörlin. "Planetary boundaries: Guiding Human Development on a Changing Planet*." Science* 347: 736.

Stephens, G. L., Li, J., Wild, M., Clayson, C. A., Loeb, N., Kato, S., L'Ecuyer, T., Stackhouse, P. W., Lebsock, M., and T. Andrews. "An Update on Earth's Energy Balance in Light of the Latest Global Observations." *Nature Geosci*. 5: 691–696.

Vitousek, P. M., Ehrlich, P. R., Ehrlich, A. H., and P. A. Matson. 1986. "Human Appropriation of the Products of Photosynthesis." *Bioscience* 36: 368–373.